\documentclass[aps,twocolumn,showpacs,showkeys,superscriptaddress]{revtex4-1}  

\usepackage{epsfig}

\begin{document} 
\title{Second-order perturbation theory  to determine the magnetic state \\ of finite size  aromatic hydrocarbons molecules}

\author{A. Valentim}
\email{alexandravalentim@cpd.ufmt.br} 
\affiliation{Centro At{\'o}mico Bariloche and Instituto Balseiro, CNEA, 8400 Bariloche, Argentina}
\affiliation{Consejo Nacional de Investigaciones Cient\'{\i}ficas y T\'ecnicas (CONICET), Argentina}
\author{D. J.  Garc\'ia}
\affiliation{Centro At{\'o}mico Bariloche and Instituto Balseiro, CNEA, 8400 Bariloche, Argentina}
\affiliation{Consejo Nacional de Investigaciones Cient\'{\i}ficas y T\'ecnicas (CONICET), Argentina}

\date{\today}   
   
\begin{abstract}  
Conjugated system have complex behaviors when increasing the number of monomers, which is  one of the reasons that makes long oligomers hard to be characterized by numerical methods.
An example of this are fused-azulene, a molecule that has been reported to displays an increasing magnetic moment with system size. 
A similar system composed  of symmetric fused-benzene rings is reported to be always no magnetic.
Instead of the empiric parametrized  Pariser-Parr-Pople (PPP) Hamiltonian, a standard model for conjugated molecules, we consider the Hubbard Hamiltonian to  explore a range of low electronic  correlation by means of perturbation theory (PT).
We show that a simple second-order perturbation treatment of electronic correlations by means of  Rayleigh-Schr\"odinger PT  allow to accurately infer about the magnetic state of these long complex $\pi$-conjugated molecules.
For fused-azulene our results supports the hypothesis that the high-spin ground state  on azulene oligomers comes from the frustrated geometry of these chains.
We validate this approach using Density Matrix Renormalization Group (DMRG) calculations. Our procedure  procedure could be helpful to describe the magnetic ground state of a  larger set of conjugated molecules.

\end{abstract}  
 
 
  
\maketitle   
  
\section{Introduction}
The study  of carbon based materials has grown to several direction spanning from engineering of electronic devices \cite{doi:10.1063/1.1767282, doi:10.1063/1.1629144, doi:10.1063/1.1767292, nature} to research on biomedical application \cite{Goenka201475,Monaco2015,Zhang2016953}.
Despite of purely organic magnetic compounds being already a reality \cite{Makarova2006vii}, the seeking for functional materials with magnetic properties such as super paramagnetism and ferromagnetism is still a goal. 
Once building blocks of such materials are molecules instead single atoms, then depending on the number of atoms to be considered, these systems can be hard to be explored computationally (for a recent review see Ref.\cite{Chiappe}). 
Reliable physical systems that has attracted great deal of attention are quasi-unidimensional molecules. Examples are fused-benzene (C$_{4n+2}$ H$_{2n+4}$) and fuzed-azulene (C$_{8n+2}$ H$_{4n+4}$) oligomers built from polycyclic aromatic hydrocarbons laterally fused.
While the first is a one-dimensional graphene stripe with zigzag edges  the second is a linear chain formed from fused azulene molecules, as so having conjugated rings with odd number of carbon atoms.

The fused-benzene series of molecules has already been synthesized up to nine monomers \cite{nonacene, ANIE:ANIE200906355} and is reported not to be magnetic.
Solid state devices of oligoacene (finite fused-benzene chain) displays a promising increasing conductivity with the number of aromatic rings \cite{doi:10.1063/1.1767282, doi:10.1063/1.1629144, doi:10.1063/1.1767292, nature}. 
Beyond nine monomers, symmetric oligoacenes (benzene chain with symmetric bond length), has been theoretically investigated  by different  many-body techniques and has also  been reported to not display magnetic polarization \cite{PhysRevB.74.045426,doi:10.1021/jp408535u,doi:10.1021/jp9015728,Simil}.
The same kind of numerical approaches predicts fused-azulene to  display a singlet-triplet transition with spin polarization of the oligomer when only a few  azulene molecules are linearly fused \cite{Chiappe, Simil, Qu}. 
Fused-azulene oligomers has also been  presented as a candidate of organic multiferroic material  \cite{Simil}, as so displaying ferroelectricity and ferromagnetic proprieties. 

\begin{figure}[h]
\setlength{\unitlength}{1.0cm}
\includegraphics[scale=0.18]{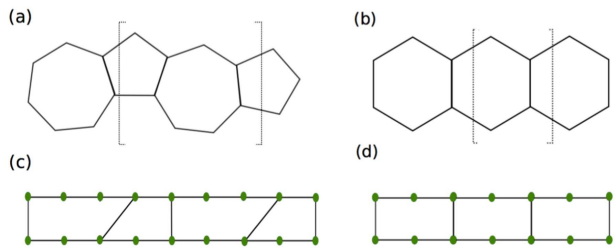}       
\caption{(color online) Schematic structure for a fused-azulene (a) and a fused-benzene molecule (b) and its respective ladder like representations (c) and (d). 
The square brackets enclosed areas represents the unit cell of each polymer structure.  The dots at ladder representations denote the positions of the carbon atoms.}
\label{fig1}
\end{figure}

These planar molecules (schematically shown at Fig.\ref{fig1}(a-b))  conventionally can be described as compose by unsaturated hydrocarbons linked by alternate double and single bonds (which are not represented at figure). 
The ladder like representation of these oligomers ( Fig.\ref{fig1}(b-c)) only differ by a twisted link in  the azulene monomer.
Within molecular orbital theory  \cite{salem1966molecular} hydrogen atoms are  neglected and the physics of the system is described  by a extended molecular orbital composed only  by  a single $p_z$ orbital on each carbon atom ($\pi$-orbitals).
At half filled band  the electronic distribution on $\pi$-orbitals allows magnetic frustration on azulene, due to the rings with odd number of carbon atoms.
For this reason we would expect  azulene molecules to have a lower energy triplet excitation than benzene molecules.

Low electronic excitations, understood as arising from delocalized electrons in the molecular orbital ($\pi$-electrons), have been appropriately described by the  empiric Pariser-Parr-Pople Hamiltonian \cite{doi:10.1063/1.1698929,TF95349FP001} for unsaturated $\pi$-conjugated organic molecules.
For symmetric fused-azulene molecules  the ground state of the  PPP model $-$ with inter-site Coulomb interaction  parametrized by the Ohno \cite{Ohno1964} formula and fixed on-site Hubbard interaction $4.69$ (in units of transfer integral between bonded sites)$-$ is singlet up to $5$-azulene. Beyond this number of monomers and up to $11$-azulene the oligomers have a triplet state \cite{Simil}.
The fuzed-azulene geometry has been also explored, in the strong correlated regime, by using  the spin-$1/2$ antiferromagnetic model Heisenberg model \cite{PhysRevB.95.224408}. The magnetic ground state moment is also observed to increase  with the monomer length.
 
In  order to explore this singlet-triplet transition, at low correlated regime, by means  of an analytic technique, here we use the Hubbard model to investigate the ground state of these conjugated systems as a function of oligomer size and electron-electron interaction.
This choice is motivated by the flexibility to change arbitrarily the parameters, then small values of electron-correlation can be treated  as a perturbation to the system.
The Hubbard Hamiltonian neglect the Coulomb repulsion between electrons of different atoms  and takes into account just the interaction between electrons in the same site (on-site Coulomb repulsion). This model   simplify the complexity of the problem while  it  retains electronic correlations and is successfully used to elicit information on correlated systems, including conjugated molecules \cite{PhysRevB.42.9088,PhysRevLett.71.1609,doi:10.1063/1.474076}.
We employ finite density matrix renormalization group (DMRG) and perturbation theory, up to second-order,  to investigate the magnetic GS of oligoacene and fused-azulene molecules as a function of electronic correlations.
The first is a powerful and complex numerical variational technique for calculating the ground state of interacting quantum systems  and  the second is a well established and analytical approach. 

Although there exist more sophisticated PT to explore the ground-state of molecules perturbatively (as M{\o}ller-Plesset \cite{doi:10.1021/ed076p170.2}), here we use a simple Rayleigh-Schr\"odinger PT to investigate transitions that take place at weak correlated regime.
In agreement with recent works \cite{doi:10.1021/acs.jctc.6b00382,PhysRevB.95.064110,1701.04800} we show that second-order PT captures the effect of electronic correlation on the magnetic state of these complex $\pi-$conjugated molecules.
We empathize that first-order PT fails to reproduce the DMRG results for the magnetic state of these oligomers however second-order PT is able to qualitative (for small oligomers) and quantitative (for larger oligomers) reproduce the numerical DMRG results.
Our results points to the frustration, arising from a combination between the geometry  of azulene and correlated $\pi$-electrons,    as the source of the magnetic polarization. 
At next section we briefly describe the model and numerical approaches used, in section III the results are presented and our final remarks are given in section IV.

\section{Model and Method}

We used the Hubbard model \cite{hubbard1963electron} to describe the electronic interactions in conjugated  systems shown at Fig.\ref{fig1}.    The model Hamiltonian contains a non-interacting part ${\hat H_0}$  and a term that incorporates the on-site electron-electron interaction $\hat{H_1}$:  
\begin{equation}                                                               
  {\hat H} = \hat{H}_0+\hat{H}_1 .     
  \label{eq1}                            
\end{equation}

The non-interacting part is a tight-binding Hamiltonian,
\begin{equation}                                                               
  \hat{H}_0= -t\sum_{<i,j>;\sigma}^{N} \left( \hat{c}^{\dagger}_{i\sigma}\hat{c}_{j\sigma}+ \hat{c}^{\dagger}_{j\sigma}\hat{c}_{i\sigma} \right) , 
  \end{equation}
where the first term describe  the kinetic energy  with a constant  hopping $t$ between nearest  neighbors atoms of carbon  (we take $t$ as the energy unit).
In the framework of $\pi-$electron theories, the operator  $\hat{c}^{\dagger}_{i\sigma}$ ( $\hat{c}_{i\sigma}$) creates (annihilate) a electron of spin $\sigma$  localized in a $\pi-$orbital at site $i$.

The interacting part of $ {\hat H}$ can be written in the form,
\begin{equation}                                                               
\hat{H}_1= U\sum_{i}^{N}\left( \hat{n}_{i\uparrow}-\frac{1}{2}\right) \left(\hat{n}_{i\downarrow}-\frac{1}{2}\right) , 
  \label{eqD}     
\end{equation}
where $U$ is the on-site Coulomb interaction and $N$ denote the number of sites, which is equal to the number of electrons in the system.
The total electron density  comported by an orbital $\pi$ is  $ \hat{n}_{i}= \hat{n}_{i\uparrow}+\hat{n}_{i\downarrow}$ and 
$\hat{n}_{i\sigma}=\hat{c}_{i\sigma}^{\dagger}\hat{c}_{i\sigma}$ is the local particle number operator for electrons of spin $\sigma$ at site $i$. At half-filling  band, the factor $1/2$   fixes the chemical potential for occupation $\langle \hat{n}_{i}\rangle=1$ on each site.

\subsection{First- and second-order perturbation theory}

In the limit of weakly correlated electrons we can treat the interaction as a small perturbation in the total energy.
The energy for the unperturbed system  is simply the energy of a tight-binding model
\begin{equation}                                                               
E^{(0)}= \langle n^{(0)}|\hat{H_0}|n^{(0)}\rangle  ,
\label{0corrE}
\end{equation}
where $|n^{(0)}\rangle$  is the ground state of the unperturbed system, which can easily be obtained once $\hat{H_0}$ is exactly diagonalizable.

Using the Rayleigh-Schr\"odinger perturbation theory \cite{sakurai2011modern} up to second-order, we can write the energy of the perturbed system as the expansion up to second-order:
\begin{equation}                                                               
E= E^{(0)}+\lambda \alpha+  \lambda^2 \beta+O(\lambda^3),
\label{0corrE}
\end{equation}
where $\lambda=U/t$,  $\lambda \alpha = \langle n^{(0)}|\hat{H_1}|n^{(0)}\rangle $ and
\begin{equation}                                                               
\lambda^{2} \beta=  \sum_{k\neq n}\frac{ |\langle n^{(0)}|\hat{H_1}|k^{(0)}\rangle |^2}{E_n^{(0)}-E_k^{(0)}},
\label{0corrE}
\end{equation}
being $|k^{(0)}\rangle$ a state of the unperturbed system. Notice that $\hat{H_1}$ is not  diagonal at the base of states  of $|k^{(0)}\rangle$. 
\newpage
Details on perturbation theory for the Hubbard model at low correlated regime are presented in \cite{SuplementaryMaterial}.
Values of $E^{(0)}$, $\alpha$ and $\beta$ to different oligoacene and fused-azulene oligomers can be find at Table \ref{table1}. 

\subsection{Density Matrix Renormalization Group}

Although the DMRG algorithm was first presented to treat an unidimensional problem \cite{ PhysRevLett.69.2863}, different low-dimensional systems can also be accurately treated with this approach \cite{Peschel:1999ae,doi:10.1080/00018730600766432}.
For the ladder like representations of oligoacene and fused-azulene in Fig.\ref{fig1} (c-d),  we can map the Hubbard Hamiltonian with first-neighbors interactions, into a unidimensional Hamiltonian with up to second-neighbors interactions, easily implemented on  a DMRG algorithm.

Oligomers described by Hamiltonian Eq.\ref{eq1}  have SU(2) symmetry so we can take advantage of the degeneracy of spin projections $M_S$, for a total spin $S$, and compute the low-lying energy of the system in a chosen subspace. The  ground state energy in $M_S$ subspace is find when $|M_S|\leq S$ and $E(S)\leq E(S+1)$, where $E(S)$ is the lowest energy belonging to a given $S$ value \cite{Lieb2004,doi:10.1063/1.1724276}.

To establish the accuracy of our DMRG simulations we have compared DMRG results for the energy at $U=0$  with the exact solution of the tight-binding Hamiltonian.
We compute the ground state-energy with different sizes of the Hilbert space, and we kept a typical cutoff  of $\rm m=800$  to a maximum of $m=1200$ states per block at final iteration.
In these conditions the energy precision is in the fourth decimal digit.
The relative energy error is comparable to the DMRG weight lost kept lower than $10^{-5}$ in the worst cases.

\section{Results and discussions}

We consider DMRG simulations and a conventional PT up to second-order to compute the energy and spin gap excitations of fused-azulene and fused-benzene-molecules with up to $74$ $\pi$-orbital ($9$ azulene and $18$ benzene monomers).
When referring to the number of $\pi$-orbital or number of atoms we are counting only carbon atoms.
We explore the Hubbard's model phase space with correlations ranging from the noninteracting limit, $U=0$, to $U=3t$ and different systems size.
We calculate the unperturbed lowest energy $E^{(0)}(M_S)$, the first-order $\alpha(M_S)$ and second-order $\beta(M_S)$  coefficients of the Rayleigh-Schr\"odinger PT for the system with spin projection $M_S=0$ and $M_S=1$.
These results, for different oligomer sizes, and the value for $U_c$ obtain by first- and second-order PT and  by DMRG  are shown  at Table \ref{table1}.
\begin{table*}[ht!]
 \centering
 \begin{tabular} { p{0.09\linewidth } p{0.04\linewidth} p{0.1\linewidth}p{0.1\linewidth}p{0.09\linewidth}p{0.1\linewidth}p{0.1\linewidth}p{0.09\linewidth}p{0.07\linewidth}p{0.07\linewidth}p{0.04\linewidth}} 
    & $N$ &$E^{(0)}(0)$ &  $\alpha(0) $ & $\beta(0)$  & $E^{(0)}(1)$  &  $\alpha(1) $ & $\beta(1)$   &  $U_c^{(1)} $ & $U_c^{(2)}$ & $U_c$   \\
    \hline
  2-azulene &18 & -24.521 & -4.449 & -0.284 & -24.016 & -4.557 & -0.271 & 4.709 & \quad- &- \\
  4-azulene &34 & -46.807 & -8.414 & -0.532 &  -46.586 &  -8.507  &-0.517  & 2.362 & \quad- &-  \\
  5-azulene &42 & -57.947 & -10.396 & -0.656 &  -57.787 & -10.485 & -0.641   & 1.796 & \quad- & 2.6 \\
  6-azulene & 50 & -69.088 & -12.378 & -0.780 &  -68.967 & -12.464 & -0. 765  & 1.409 & 2.376 & 2.0 \\
  7-azulene & 58 & -80.228 & -14.360 & -0.904 &  -80.133 & -14.443 & -0. 889  & 1.134 & 1.593 & 1.8 \\
  9-azulene &74 & -102.509 & -18.324 & -1.153 &   -102.447 & -18.404 & -1.137   & 0.779 &0.965 & 1.0 \\
  4-acene &18 & -24.930 & -4.500 & -0.294 & -24.340 & -4.604 & -0.275 & 5.636 &\quad- &- \\
  10-acene &42 & -58.601 & -10.500 & -0.696 & -58.451 & -10.559 & -0.658 & 2.518 &\quad- &- \\
  18-acene &74 & -103.492 & -18.500 & -1.242 & -103.439 & -18.537 & -1.183 & 1.415 &\quad- &- \\
    \hline
  \end{tabular}
\caption{Non-interacting lowest energy $E^{(0)}(M_S)$, first- $\alpha(M_S)$ and second-order $\beta(M_S)$  coefficients of perturbation theory and critical value of electronic correlation $U_c$ obtain by FOPT,  SOPT, and  DMRG calculations respectively  (all  in units of $t$) for different number of monomers ($N$ atoms) in  oligoacene and fused-azulene molecules.
\label{table1}}
\end{table*}

We find that by using fist-order PT both systems undergoes a singlet-triplet state transition in a $U_c$ which depend of the oligomer size.
However by means of second-order PT we qualitatively reproduce the numerical predictions, that oligoacene always has a singlet ground state while fused-azulene presents a increasing ground state with the  chain length.
DMRG calculations validates this analysis, with the agreement between different approaches increasing with the size of the system.
We take $U/t\lesssim2$ as the upper limit of validity for the results obtain from perturbation theory, as for larger $U/t$ values $H_1$ becomes the most relevant term in the Hamiltonian \cite{khomskii2010basic,essler2005one}. 

\begin{figure}[h!]
\setlength{\unitlength}{1.0cm}
\vspace{-0.4cm}
\includegraphics[scale=0.32]{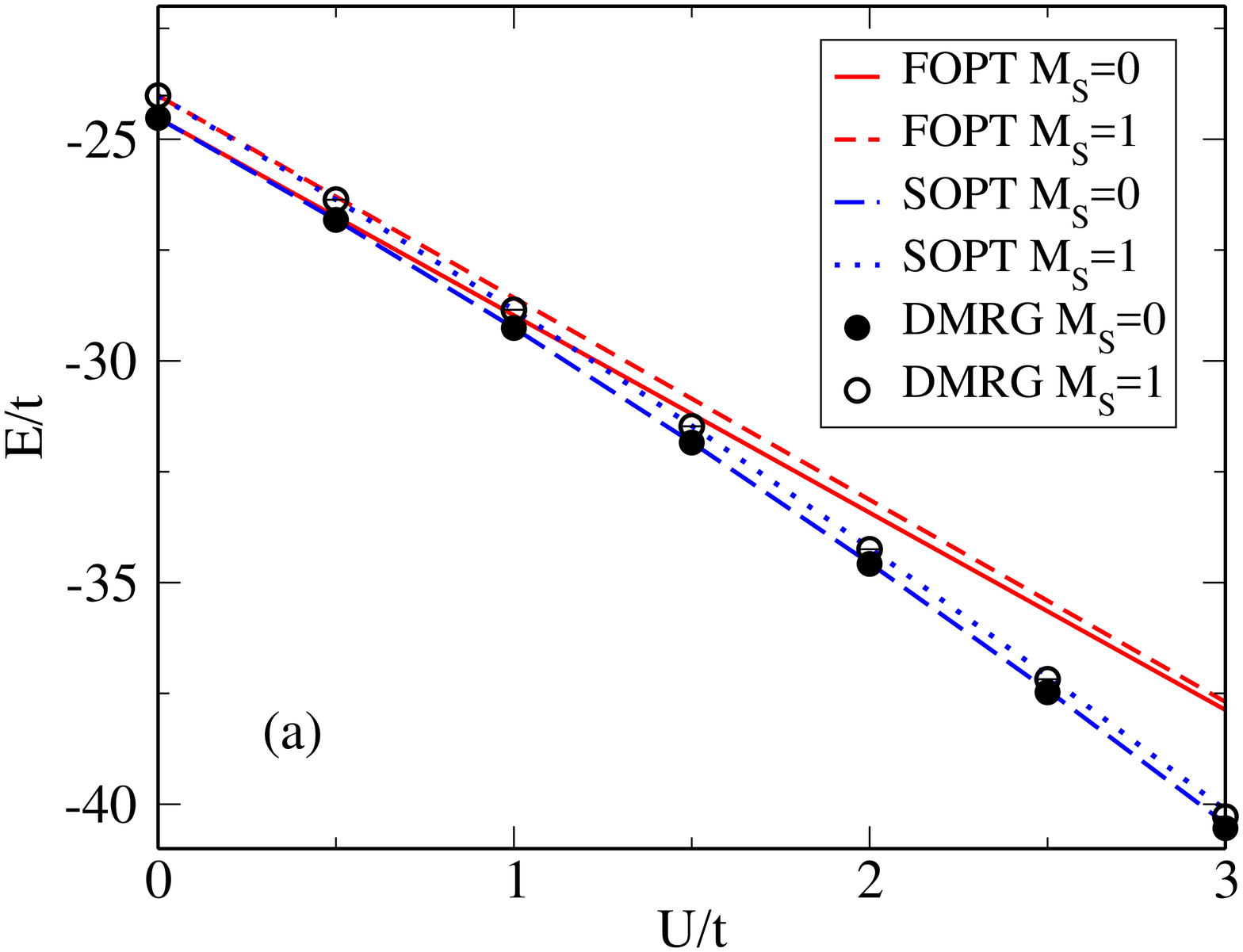} 
\vspace{-0.4cm}
\includegraphics[scale=0.32]{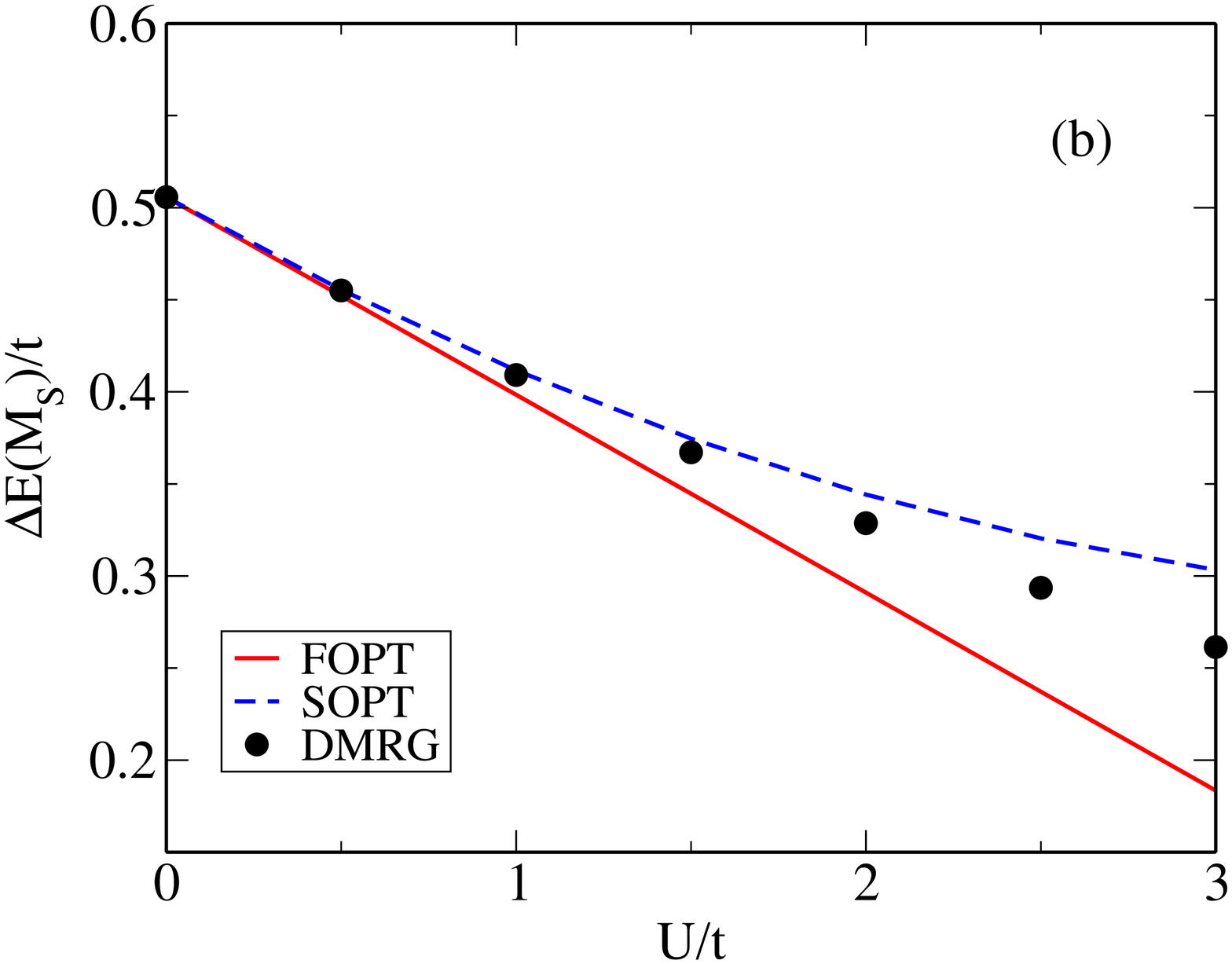}  
\caption{ (color online) Panel (a) shows low-lying energies $\rm E(M_S=0)$  and $\rm E(M_S=1)$,  and panel (b) shows the singlet-triplet spin gap  $\rm \Delta E(M_{S})$, both as a function of $\rm U$ for a fused-azulene molecule with $2$ monomers ($18$ atoms).
 The DMRG data has a error bar which is smaller than the symbols size.} 
\label{Graph1}
\end{figure}

To estimate $U_c$ we calculate the singlet-triplet spin gap defined as $\Delta E(M_S)= E(M_S=1)-E(M_S=0)$ and define $U_{c}$ as the value where this gap closes $\Delta E(M_S)=0$.
This analysis is illustrated in  Fig.\ref{Graph1}(a-b) for two linearly fused $2$-azulene molecule ( $18$ atoms).
Panel (a) shows  the energy computed on the $M_S=0$ and $M_S=1$ spaces.
We observe that the curve  for $E(M_S=0)$ lies slightly below $ E(M_S=1)$ curve, what  indicates  a singlet state with a finite gap between the ground state energy,  $E(M_S=0)$, and the energy of the first magnetic excited state,  $E(M_S=1)$.
 As expected, DMRG is the lowest energy and second-order PT (SOPT) data is closer to DMRG results than first-order PT (FOPT).
At  panel (b) we plot the singlet-triplet spin gap as a function of electronic correlations.  A finite gap indicating a singlet state is found in agreement with Refs.\cite{Chiappe, Simil, Qu}.
FOPT results points out to  a transition from singlet to triplet ground state for values of $U_c$ above the validity of the approximation (not displayed at the plot).

\begin{figure}[h!]
\setlength{\unitlength}{1.0cm}
\includegraphics[scale=0.39]{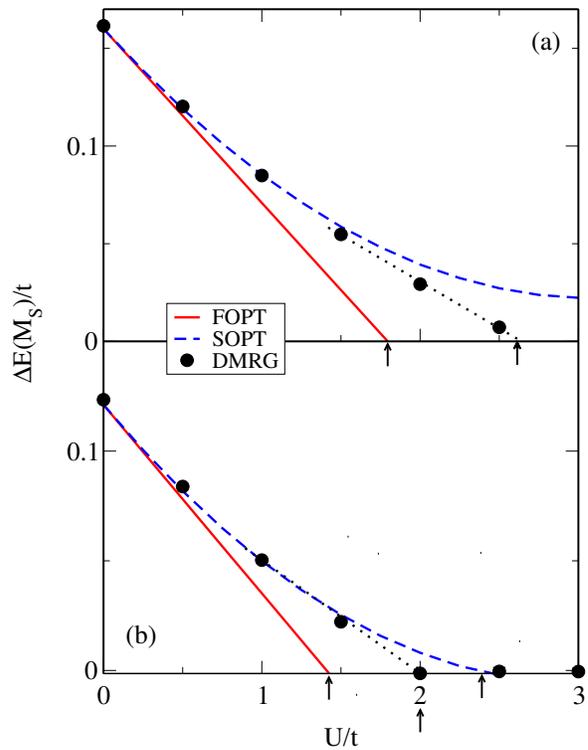}   
\caption{ (color online) Singlet-triplet spin gap  $\rm \Delta E(M_{S})$ as a function of $ \rm U/t$ for (a) $5$-azulene and (b) $6$-azulene oligomers with  $42$ and $50$ atoms respectively.  Arrows points the values of $\rm U_c$ estimated with each approach. 
The DMRG data has a error bar which is smaller than the symbols size.}
\label{Graph2}
\end{figure}

In  Fig.\ref{Graph2}, we plot the singlet-triplet spin gap as a function of electronic correlations for two different sizes of fused-azulene molecules. 
FOPT and DMRG calculations  displays a finite gap in both cases, for small values of $U/t$, indicating a singlet state.
SOPT estimate $U_c$  below $2t$ when the chain is larger than $6$-azulene. 
In Table \ref{table1} we show that the agreement of the value of $U_c$ between the different approaches increases with the number of monomers.
These results on fused-azulenes could lead to the incorrect assumption that first-order treatment  of electronic correlations is enough to describe these molecules.
We show that for fused-benzene FOPT incorrectly predicts a singlet-triplet transition while SOPT   correctly  predicts that the ground state for fused-benzene is always a singlet.

\begin{figure}[h!]
\setlength{\unitlength}{1.0cm}
\includegraphics[scale=0.32]{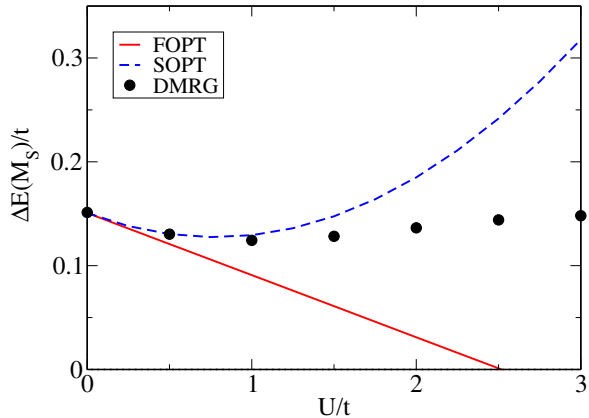}  
\caption{ (color online) Singlet-triplet spin gap  $\rm \Delta E(M_{S})$ as a function of $ \rm U/t$ for $10$-benzene oligomer with $42$ atoms. The DMRG data has a error bar which is smaller than the symbols size.}
\label{Graph3}
\end{figure}

At Fig.\ref{Graph3} we repeat the analysis of the singlet-triplet spin gap for ten fused monomers of benzene ($10$-acene with $42$ atoms).  
DMRG simulations and SOPT calculus exhibit a significant spin gap while using FOPT the spin gap goes to zero. 
Second-order PT predicts a singlet ground state for all values of $U/t$ in agreement with DMRG.
We checked these results up to $18$ fused monomers (see Table \ref{table1}) and we can reproduce the expected behavior of spin excitations of oligocene chains, i.e. a finite spin gap which indicates that the ground state of the system is a singlet independently of the electronic correlation magnitude and system size \cite{Simil}. 

\begin{figure}[h!]
\setlength{\unitlength}{1.0cm}
\includegraphics[scale=0.34]{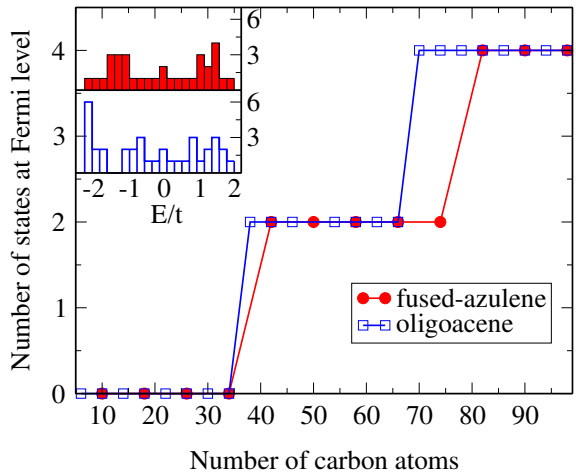}  
\caption{ (color online)  Accumulation of the number of states near the Fermi level for oligoacene and azulene molecules as a function of oligomer size. The number of states is computed within a range of $0.1t$ around  the Fermi level for the tight-binding model. The inset displays the number of states over the whole energy spectrum for  fused-azulene and oligoacene chains with $42$ atoms.}
\label{Graph4}
\end{figure}

The possibility for ferromagnetism in odd-membered rings oligomers (fused-azulene geometry) could be related to flat-band magnetism \cite{PhysRevLett.105.266403,PhysRevB.68.174419}, a ferromagnetic behavior due to the accumulation of degenerated states at Femi level.
As shown at Fig.\ref{Graph4} both $n$-acene and $n$-azulene accumulates states in the Fermi level, but only fused-azulene shows a transition to a high spin state, ruling out flat-band as the source for the magnetism on azulene oligomers. 
Geometric frustration on $n$-azulene increases as the chain length increases (more twisted links as in Fig.\ref{fig1}c) and the Coulomb repulsion needed for a $S=0$ to $S=1$ transition diminish. 
These results points to frustration (absent on oligoacene) as the source of the magnetic transition.

DMRG results on antiferromagnetic spin-$1/2$ Heisenberg Hamiltonian  for both azulene and benzene oligomers  shows that the magnetic state does not require the existence of flat energy bands \cite{Simil}.
Configuration interaction on PPP model  also reports a $S=0$ ground state for $2$-azulene and $3$-azulene with a decreasing of the singlet-triplet spin gap as a function of the chain length. 
They argue this would indicate a tendency to a magnetic ground state, however  simulations on bigger oligomer sizes are need   in order to confirm the prediction by   DMRG \cite{Simil}.
The authors also conjecture  the increasing frustration percentage (number of frustrated links, $n$, divided by the number of non-frustrated links, $9n+1$)  as the origin of a possible higher ground state for long azulene oligomers \cite{Chiappe}.
In agreement with Ref.\cite{Simil} and our results, density functional theory methods also reports  a triplet state for $6$-azulene \cite{Qu}.
In summary, second-order PT allowed us to explore the low correlation  regime ($U/t\lesssim 2$) and find the same magnetic transitions shown by more sophisticated and computationally demanding numerical methods.
In the specific case of fused-azulene our second-order PT results show that a magnetic transition in this weak interaction regime happens for oligomers with more than four monomers.

\vspace{1em}
\section{Conclusion}

We employ Rayleigh-Schr\"odinger perturbation theory and DMRG technique to study low-lying spin excitations for two geometrically related conjugated systems.
We investigate the ground state of these conjugated molecules by using the Hubbard model with weak on-site coulomb interaction, ranging from the non-interacting limit to $U/t=3$, and different monomer length.
We show that second-order perturbation theory is a reliable tool to characterize magnetic transitions on complex molecules, using as a non trivial example fused-benzene and fused-azulene oligomers. 
Our results support that for  $n$-azulene there is a  magnetic transition from singlet to triplet as the size the system increases. 
This transition is not related to flat-band magnetism, instead, the comparison with oligoacene suggest that the origin of magnetism in fused-azulene  oligomers comes from the frustrated geometry of these chains. 
For fused-azulene the value of $U_c$ in which a transition from a singlet to a triplet state take place decrease  with the size of the system, being lower than $2t$ for oligomers bigger than $4$-azulene.
The increasing  agreement between SOPT  and DMRG simulations,  for longer oligomers, makes this approach a useful tool to the investigation of conjugated organic molecules, at weak correlated regime, to  system-sizes that would be impossible by means of numeric techniques.

\section{Supplementary material} 
See supplementary material (1) for a detailed  derivation of perturbation theory for the Hubbard model at low correlated  regime   and supplementary material (2) for  a Maxima \cite{Maxima} script to compute  the fist- and second-order quantities presented in Table \ref{table1}.

\section{Acknowledgments} 
We acknowledge a fruitful conversation with Suchi Guha.
This work has been supported by CONICET and PICT  2012/1069.

%
\end{document}